\documentclass[%
 reprint,
superscriptaddress,
 amsmath,amssymb,
 aps,
]{revtex4-2}

\usepackage{graphicx}
\usepackage{dcolumn}
\usepackage{bm}

\usepackage[colorlinks=true]{hyperref}
\usepackage{upgreek}
\usepackage{physics}
\usepackage{xcolor}
\definecolor{BrewerSetRed}{RGB}{228,26,28}
\definecolor{BrewerSetBlue}{RGB}{55,126,184}
\definecolor{BrewerSetGreen}{RGB}{77,175,74}
\definecolor{BrewerSetPurple}{RGB}{152,78,163}
\definecolor{BrewerSetOrange}{RGB}{255,127,0}
\definecolor{BrewerSetYellow}{RGB}{236,176,23}
\definecolor{BrewerSetBrown}{RGB}{166,86,40}
\definecolor{BrewerSetPink}{RGB}{247,129,191}
\definecolor{BrewerSetGray}{RGB}{153,153,153}

\usepackage{tikz}
\usetikzlibrary{shapes.geometric,positioning}

\DeclareRobustCommand\Circle[1]{
    \tikz[baseline=-0.6ex]
    {
        \draw[very thick, #1] ({2em/2}, 0) circle [radius = 0.25em];
    }
}

\begin{document}

\preprint{APS/123-QED}

\title{Higher-order Skin Effect through a Hermitian-non-Hermitian Correspondence \\ and Its Observation in an Acoustic Kagome Lattice}

\author{Jia-Xin Zhong}
\thanks{These authors contributed equally to this work.}%
\affiliation{Graduate Program in Acoustics, The Pennsylvania State University, University Park, PA 16802, USA}
\author{Pedro Fittipaldi de Castro}
\thanks{These authors contributed equally to this work.}%
\affiliation{Department of Physics, Emory University, Atlanta, Georgia 30322, USA}
\author{Tianhong Lu}
\affiliation{Department of Physics, Emory University, Atlanta, Georgia 30322, USA}
\author{Jeewoo Kim}
\affiliation{Graduate Program in Acoustics, The Pennsylvania State University, University Park, PA 16802, USA}
\author{Mourad Oudich}
\affiliation{Graduate Program in Acoustics, The Pennsylvania State University, University Park, PA 16802, USA}
\affiliation{Université de Lorraine, CNRS, Institut Jean Lamour, F-54000 Nancy, France}
\author{Jun Ji}
\affiliation{Graduate Program in Acoustics, The Pennsylvania State University, University Park, PA 16802, USA}
\author{Li Shi}
\affiliation{Key Laboratory of Modern Acoustics and Institute of Acoustics, Nanjing University, Nanjing 210093, China}
\author{Kai Chen}
\affiliation{Key Laboratory of Modern Acoustics and Institute of Acoustics, Nanjing University, Nanjing 210093, China}
\author{Jing Lu}
\email{lujing@nju.edu.cn}
\affiliation{Key Laboratory of Modern Acoustics and Institute of Acoustics, Nanjing University, Nanjing 210093, China}
\author{Yun Jing}
\email{yqj5201@psu.edu}
\affiliation{Graduate Program in Acoustics, The Pennsylvania State University, University Park, PA 16802, USA}
\author{Wladimir A. Benalcazar}
\email{benalcazar@emory.edu}
\affiliation{Department of Physics, Emory University, Atlanta, Georgia 30322, USA}

\date{\today}

\begin{abstract}
    The non-Hermitian skin effect (NHSE) is a distinctive topological phenomenon observed in non-Hermitian systems.
    Recently, there has been considerable interest in exploring higher-order NHSE occurrences in two and three dimensions.
    In such systems, topological edge states collapse into a corner while bulk states remain delocalized.
    Through a Hermitian-non-Hermitian correspondence, this study predicts and experimentally observes the higher-order NHSE in an acoustic Kagome lattice possessing nonreciprocal hoppings.
    By rotating the frequency spectrum and employing complex-frequency excitation techniques, we observe the localization of acoustic energy towards a corner of the lattice in the topologically nontrivial phase, even when the source is located far from that corner.
    In contrast, the acoustic energy spreads out when excited at the frequencies hosting the bulk states.
    These observations are unequivocal evidence of the higher-order NHSE.
\end{abstract}

\maketitle


\section{Introduction}

Along with quantized bulk signatures, topological boundary states constitute a hallmark of topological phases of matter.
The bulk and boundary properties in these phases are related by a \emph{bulk-boundary correspondence}, first identified in Hermitian systems and then explored in non-Hermitian systems that exhibit phenomena such as the non-Hermitian skin effect (NHSE)~\cite{Kunst2018BiorthogonalBulkBoundaryCorrespondence, Yang2020NonHermitianBulkBoundaryCorrespondence, Xiao2020NonHermitianBulkBoundary}.
In Hermitian systems, topological boundary states are characterized by topological invariants of the bulk states within the system.
A $d$-dimensional system with $L^{d}$ degrees of freedom in an $n$th order topological phase has $\mathcal{O}(L^{d-n})$ boundary states~\cite{Kawabata2020HigherorderNonHermitianSkin, Benalcazar2017QuantizedElectricMultipole, Schindler2018HigherorderTopologicalInsulators}.
Recent studies have generalized the NHSE in 1D to the high-order NHSE in higher dimensions \cite{Lee2019HybridHigherOrderSkinTopological, Okugawa2020SecondorderTopologicalNonHermitian, Kawabata2020HigherorderNonHermitianSkin, Fu2021NonHermitianSecondorderSkin, Zhang2022ReviewNonHermitianSkin, Lin2023TopologicalNonHermitianSkin}, showing that systems with an $n$th order NHSE have $\mathcal{O}(L^{d-n+1})$ skin modes \cite{Kawabata2020HigherorderNonHermitianSkin, Okugawa2020SecondorderTopologicalNonHermitian,Lu2023NonHermitianTopologicalPhononic}.
However, these states occupy a boundary of codimension $d-n$, resulting in a collapse of $\mathcal{O}(L)$ boundary states in the $d-n$ dimensional boundary.

Recently, many experimental observations of the NHSE on various physical platforms have been reported.
The first-order NHSE was initially demonstrated in 1D photonic lattices \cite{Weidemann2020TopologicalFunnelingLight}, quantum systems \cite{Xiao2020NonHermitianBulkBoundary,Liang2022DynamicSignaturesNonHermitian}, topolectrical circuits \cite{Helbig2020GeneralizedBulkBoundary}, mechanical systems \cite{Ghatak2020ObservationNonHermitianTopology}, and acoustic lattices \cite{Zhang2021AcousticNonHermitianSkin, Gu2022TransientNonHermitianSkin}.
Higher-order NHSE has also been observed in 2D and 3D systems recently, including topolectrical circuits \cite{Zou2021ObservationHybridHigherorder, Liu2023ExperimentalObservationNonHermitian}, acoustic lattices \cite{Zhang2021ObservationHigherorderNonHermitian, Zhang2024ConstructionObservationFlexibly, Gao2024ControllingAcousticNonHermitian}, and active particles \cite{Palacios2021GuidedAccumulationActive}.
Nevertheless, all of these experimental observations occurred in square or cubic lattices whose Hamiltonians are separable; that is, they can be written as $h(k_x,k_y)=h(k_x)\otimes I +I\otimes h(k_y)$, where $I$ is the identity matrix \cite{Wang2021ConstructingHigherorderTopological}.
Consequently, their eigenstates take the simple form $\psi(k_x,k_y)=\psi(k_x)\otimes\psi(k_y)$ with corresponding energies $E(k_x)+E(k_y)$, where $\psi(k_i)$, for $i=x,y$, are the eigenstates to the 1D problem, $h(k_i)\psi(k_i)=E(k_i)\psi(k_i)$.
In this sense, they are straightforward extensions of the 1D Hatano-Nelson and/or non-Hermitian Su-Schrieffer-Heeger models to higher dimensions.
If the higher-order NHSE is a general feature of non-Hermitian lattices with higher-order topology, the skin modes should also arise in other (non-separable) lattices protected by crystalline symmetries.
However, a systematic approach to thoroughly classify and understand the topological mechanisms that protect these states is still lacking.

Here, we use the Hermitian-non-Hermitian correspondence~\cite{Kawabata2020HigherorderNonHermitianSkin}, by which every Hermitian Hamiltonian protected by chiral symmetry has a corresponding non-Hermitian Hamiltonian carrying the same topological information.
As a specific example, we demonstrate that the non-Hermitian dimerized Kagome lattice with nonreciprocal nearest-neighbor hoppings must exhibit a higher-order NHSE.
The parent Hermitian Hamiltonian of this Kagome lattice is the honeycomb lattice with Kekul\'e distortion, otherwise known as ``breathing honeycomb'' lattice, which has been shown to exhibit a higher-order topological insulator (HOTI) phase protected by chiral symmetry~\cite{Noh2018TopologicalProtectionPhotonic}.
Unlike previous models~\cite{Wang2023NonHermitianTopologicalPhases}, the particular non-Hermitian Kagome lattice resulting from this correspondence does not require long-range hoppings, making its realization far more feasible.
We then experimentally verify this effect in a non-Hermitian acoustic Kagome lattice composed of cavities interconnected in a nonreciprocal (unidirectional) manner.

Topological acoustics is a versatile platform for realizing various types of topological phases, including those exhibiting the NHSE \cite{Zhang2021AcousticNonHermitianSkin, Zhang2021ObservationHigherorderNonHermitian}.
While the NHSE can be realized in passive acoustic lattices through intentionally induced losses, it is often significantly obscured by the substantial loss \cite{Zhang2021ObservationHigherorderNonHermitian, Gu2022TransientNonHermitianSkin, Wang2023ExperimentalRealizationGeometrydependenta}.
Reports have shown that it is notably more flexible and straightforward to observe the NHSE in an active acoustic lattice by introducing gain \cite{Zhang2021AcousticNonHermitianSkin,  Wang2023ExtendedTopologicalMode, Wang2022NonHermitianMorphingTopological}.
In these existing active acoustic lattices, reciprocal hoppings are provided by physically connected passive components such as tubes, while nonreciprocal hoppings are realized by pairs of sources (e.g., loudspeakers) and detectors (e.g., microphones).
In contrast to these acoustic lattices, our implemented lattice features sites interconnected solely by nearest-neighbor unidirectional hopping, thereby simplifying its implementation.
To more clearly observe the NHSE, we first intentionally rotate the complex energy spectrum by implementing complex hoppings to minimize the loss of skin states while maintaining the system's stability.
Additionally, we implement the complex-frequency excitation (CFE) technique, which was recently reported to provide virtual gain to compensate for the loss in a system without causing stability issues \cite{Li2020VirtualParityTimeSymmetry, Kim2022BoundsLightScattering, Kim2023LossCompensationSuperresolution, Guan2023OvercomingLossesSuperlenses, Guan2024CompensatingLossesPolariton, Zeng2024SynthesizedComplexfrequencyExcitation}.
The CFE technique has also been introduced to augment the observation of the NHSE in non-Hermitian acoustic lattices, where much clearer energy localization can be observed due to the compensation for the background loss \cite{Gu2022TransientNonHermitianSkin, Gao2024ControllingAcousticNonHermitian}.
By combining these techniques, we observe a significant confinement of acoustic energy at a specific corner in the topologically nontrivial phase when sites at the edges are excited---even if the excitation site is located far from that corner.
In contrast, the acoustic energy spreads out throughout the bulk when bulk states are excited.
Both observations present evidence of a higher-order NHSE, the first one observed in a lattice with non-separable 2D Bloch Hamiltonian.

\section{Chiral-symmetric HOTI and the higher-order NHSE}

We may find systems supporting the higher-order NHSE protected by $C_n$ symmetry---the symmetry that leaves a crystal invariant upon rotation by $2\pi/n$ rad---by noting that, for each Hermitian HOTI $h(\bm{k})$ protected by chiral symmetry,
\begin{equation}
    \Gamma h(\bm{k}) \Gamma^{-1} = -h(\bm{k}), \label{Chiral}
\end{equation}
as well as $C_{2n}$ symmetry,
\begin{equation}
    r_{2n} h(\bm{k}) r_{2n}^{\dagger} = h(R_{2n}\bm{k}), \label{Cn}
\end{equation}
there is a corresponding non-Hermitian Hamiltonian with the same topological information as $h(\bm{k})$, obeying $C_n$ symmetry. In Eq.~\eqref{Chiral}, $\Gamma$ represents the chiral operator while, in Eq.~\eqref{Cn}, $r_{2n}$ is the unitary $C_{2n}$ rotation operator and $R_{2n}$ is the rotation matrix of the momentum vector $\bm{k}$ by an angle $\pi/n$, both satisfying $(r_{2n})^{2n}= I_{2N\times 2N}$ and $(R_{2n})^{2n}=I_{2\times 2}$, where the unit cell of the crystal has $2N$ degrees of freedom.

For Hermitian systems, chiral symmetry \eqref{Chiral} is equivalent to sublattice symmetry.
Consequently, one can always choose a basis where the chiral operator is $\Gamma = \sigma_z \otimes I_{N}$, making the Bloch Hamiltonian take the off-diagonal form
\begin{equation}
    h(\bm{k}) = \begin{pmatrix}
        0                   & q(\bm{k}) \\
        q^{\dagger}(\bm{k}) & 0
    \end{pmatrix}, \label{eq:Hamiltonian}
\end{equation}
where $q(\bm{k})$ is generally a non-Hermitian $N \times N$ matrix. In the basis in which the Hamiltoninan in Eq.~\eqref{eq:Hamiltonian} is written, $r_{2n}$ is block diagonal,
\begin{equation}
    r_{2n} = \begin{pmatrix}
        r_n & 0   \\
        0   & r_n
    \end{pmatrix}, \quad \mathrm{where} \quad r_nq(\bm{k})r_n^{\dagger} = q(R_n\bm{k}). \label{Small}
\end{equation}
Reference \cite{Benalcazar2019QuantizationFractionalCorner} has shown that under the symmetry conditions \eqref{Chiral} and \eqref{Cn}, Hamiltonians of the form in Eq.~\eqref{eq:Hamiltonian} having corner-induced filling anomalies host corner-localized zero-energy states.

Here, we observe that all the information of $h(\bm{k})$ is contained in the block $q(\bm{k})$, which possesses $C_n$ symmetry (see Eq.~\eqref{Small}).
The Hermitian-non-Hermitian correspondence involves interpreting $q(\bm{k})$ as a non-Hermitian Hamiltonian in its own right.
Since both $h(\bm{k})$ and $q(\bm{k})$ carry the same topological information, Hermitian Hamiltonians $h(\bm{k})$ describing chiral symmetric HOTIs with higher-order boundary states will have a non-Hermitian counterpart $q(\bm{k})$ with nontrivial higher-order topology and higher-order boundary states.
One can visualize the connection between chiral zero-energy modes in $h(\bm{k})$ and skin modes in $q(\bm{k})$ by noting that the chiral zero-energy modes of a bipartite Hamiltonian, such as Eq. \eqref{eq:Hamiltonian}, are supported on one sublattice and, for that reason, are also
exponentially localized eigenstates of $q(\bm{k})$ or $q^{\dagger}(\bm{k})$ with zero energy.
See Supplementary Note S1 for a more detailed discussion in one dimension.

In the following, we apply the Hermitian-non-Hermitian correspondence to the 2D breathing honeycomb lattice~\cite{Noh2018TopologicalProtectionPhotonic}, a gapped Hermitian lattice with zero-energy corner states protected by chiral and $C_6$ (and thus $C_3$) symmetries, to define a non-Hermitian Kagome lattice supporting a second-order NHSE protected by $C_3$ symmetry.
It should be emphasized that the theoretical framework here is generic and, therefore, can be applied to other lattices.

\subsection{Correspondence of Hermitian breathing honeycomb and non-Hermitian Kagome models}

\begin{figure}[!htb]
    \centering
    \includegraphics[width = 0.45\textwidth]{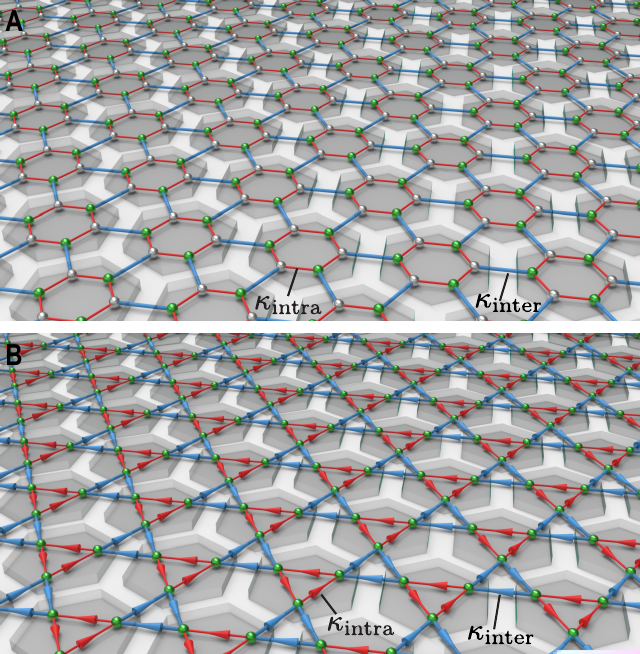}
    \caption{Hermitian HOTI and its non-Hermitian counterpart.
        (\textbf{A}) The Hermitian breathing honeycomb lattice, where all hoppings are reciprocal.
        (\textbf{B}) The non-Hermitian Kagome lattice, where all hoppings are unidirectional.
        The intra- (red) and inter-cell (blue) hoppings are denoted by $\kappa_\mathrm{intra}$ and $\kappa_\mathrm{inter}$, respectively.
        For simplicity, their ratio is defined as $\kappa = \kappa_\mathrm{intra}/\kappa_\mathrm{inter}$.
    }
    \label{TheoryFig1}
\end{figure}

The breathing honeycomb model consists of a 2D triangular Bravais lattice where each unit cell has six sites arranged in a hexagon (see Fig.~\ref{TheoryFig1}A).
The tight-binding model (TBM) contains nearest-neighbor hopping terms with strength $\kappa_{\mathrm{intra}}$ between sites within the same unit cell and inter-cell nearest-neighbor hoppings with strength $\kappa_{\mathrm{inter}}$, giving rise to six energy bands.
Due to the model's chiral (sublattice) symmetry, the three lower bands lie below the zero energy level, while the upper bands are reflections of the first three about $E=0$.
In Fig.~\ref{TheoryFig1}A, we distinguish the two sublattices by their color---white or green.

In addition to chiral symmetry, the breathing honeycomb model possesses time-reversal symmetry (TRS) and belongs to class BDI in the 10-fold classification of Hermitian Hamiltonians \cite{Altland1997NonstandardSymmetryClasses}.
This class does not exhibit nontrivial topological phases at first order in two dimensions.
However, the additional $C_6$ symmetry protects second-order topological zero energy states at specific corners of the lattice when $\kappa \equiv \kappa_{\mathrm{intra}}/\kappa_{\mathrm{inter}}<1$~\cite{Noh2018TopologicalProtectionPhotonic}.

The presence of chiral  symmetry allows us to express the breathing honeycomb's $6 \times 6$ Bloch Hamiltonian in the off-diagonal form Eq.~\eqref{eq:Hamiltonian}, with the
$3 \times 3$ non-Hermitian block $q(\bm{k})$ given by
\begin{equation}
    q(\bm{k}) = \begin{pmatrix}
        0                                                  & \kappa_\mathrm{intra}                               & \kappa_\mathrm{inter} e^{i{\bm{k}}\cdot{\bm{ a}}_1} \\
        \kappa_\mathrm{inter}e^{i{\bm{k}}\cdot{\bm{ a}}_2} & 0                                                   & \kappa_\mathrm{intra}                               \\
        \kappa_\mathrm{intra}                              & \kappa_\mathrm{inter} e^{i{\bm{k}}\cdot{\bm{ a}}_3} & 0 \label{KagomeHamiltonian}
    \end{pmatrix}.
\end{equation}
Here, ${\bm{a}}_1 = (1,0), {\bm{ a}}_2 = -(1/2,\sqrt{3}/2), {\bm{ a}}_3 = (-1/2,\sqrt{3}/2)$ are primitive lattice vectors connecting neighboring unit cells.

When taken as a non-Hermitian Hamiltonian, the matrix $q(\bm{k})$ in Eq.~(\ref{KagomeHamiltonian}) describes a non-Hermitian Bloch Hamiltonian on a Kagome lattice, consisting of a triangular lattice of unit cells, each having three sites, and where the nearest-neighbor hoppings are unidirectional, having strengths $\kappa_{\mathrm{intra}}$ (for intra-cell hoppings) and $\kappa_{\mathrm{inter}}$ (for inter-cell hoppings) as shown in Fig.~\ref{TheoryFig1}B.
The Bloch Hamiltonian in Eq.~(\ref{KagomeHamiltonian}) satisfies the $C_3$ symmetry condition
\begin{equation}
    r_3 q(\bm{k}) r_3^{\dagger} = q(R_3 \bm{k}), \label{rotationsymmetry}
\end{equation}
where $R_3$ is the rotation matrix by an angle $2\pi/3$, and
$    r_3 = \begin{pmatrix}
        0 & 1 & 0 \\
        0 & 0 & 1 \\
        1 & 0 & 0
    \end{pmatrix}
$
is the $C_3$ rotation operator.
Additionally, $q(\bm{k})$ is time-reversal symmetric:
%
$\Theta q(\bm{k})\Theta^{-1} = q(-\bm{k}),$
%
where $\Theta$ represents complex conjugation for our spinless model.
However, it is important to note that for non-Hermitian Hamiltonians, there is a ramification of TRS into an additional symmetry condition, $\Theta q^{\dagger}(\bm{k})\Theta^{-1} = q(-\bm{k})$~\cite{Kawabata2019SymmetryTopologyNonHermitian}, and that $q(\bm{k})$ violates this condition.

Analogous to its Hermitian counterpart, $q(\bm{k})$ falls into symmetry class AI in the 38-fold classification for non-Hermitian Hamiltonians~\cite{Kawabata2019SymmetryTopologyNonHermitian}.
This class has a winding number of zero in 2D, indicating that our system does not host the first-order NHSE.
However, the presence of crystalline symmetries, specifically $C_3$ in this case, enriches the 38-fold classification, resulting in higher-order topological phases that exhibit the higher-order NHSE.



\subsection{Topological phases and bulk topological invariants}

\begin{figure*}
    \centering
    \includegraphics[width=0.9\textwidth]{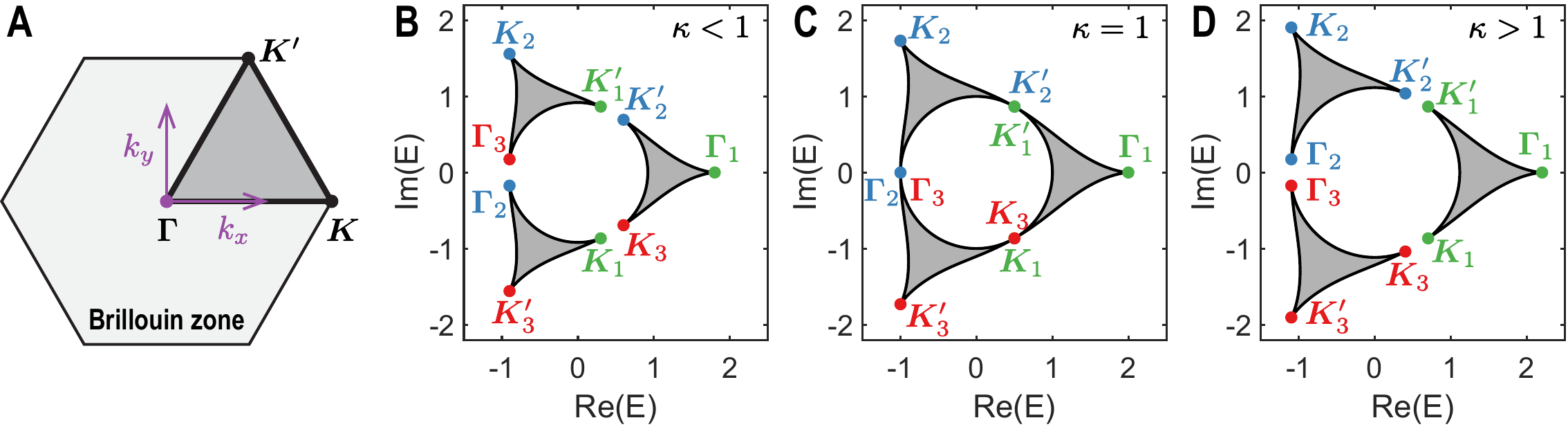}
    \caption{
    (\textbf{A}) Brillouin zone for the non-Hermitian Kagome lattice showing the locations of the HSPs $\bm{\Gamma}=(0,0)$, $\bm{K}=(4\pi/3,0)$, and $\bm{K}'=(2\pi/3,2\pi/\sqrt{3})$.
    (\textbf{B}--\textbf{D}) Representations of $C_3$ symmetry  for bulk states at the Brillouin zone points $\bm{K}$, $\bm{K}'$, and $\bm{\Gamma}$ in the (\textbf{B}) topological phase,  (\textbf{C}) transition point, and (\textbf{D}) trivial phase.
    The respective parameters are $\kappa=0.8$, $\kappa=1$, and $\kappa=1.2$, with $\kappa \equiv \kappa_\mathrm{intra} / \kappa_\mathrm{inter}$ and $\kappa_\mathrm{inter} = 1$.
    Green (\Circle{fill, color=BrewerSetGreen}), blue (\Circle{fill, color=BrewerSetBlue}), and red (\Circle{fill, color=BrewerSetRed}) dots represent the values $\Pi^{(3)}_1=1$, $\Pi^{(3)}_2={e}^{2\pi {i}/3}$, and $\Pi^{(3)}_3=e^{-2 \pi {i}/3}$, respectively.
    }
    \label{HighSymmetryPoints}
\end{figure*}

We can distinguish the topological phases of non-Hermitian Hamiltonians protected by $C_3$ symmetry through the irreducible representations that the eigenstates take at high-symmetry points (HSPs) of the Brillouin zone $\bm{\Pi}=\bm{K},\bm{K}',\bm{\Gamma}$ (Fig.~\ref{HighSymmetryPoints}A).
These three HSPs obey $R_3\bm{\Pi} = \bm{\Pi}$ modulo a reciprocal lattice vector, and $C_3$ symmetry (Eq.~\eqref{rotationsymmetry}) implies $[r_3,q(\bm{\Pi})]=0$. Therefore, the eigenstates of the energy band $m$ at the HSPs, which satisfy $q(\bm{\Pi})\ket{\varepsilon_m(\bm{\Pi})}=\varepsilon_m(\bm{\Pi})\ket{\varepsilon_m(\bm{\Pi})}$, are also the eigenstates of the rotation operator, $r_3\ket{\varepsilon_m(\bm{\Pi})}=\Pi_p^{(3)}\ket{\varepsilon_m(\bm{\Pi})}$.
For spinless systems, the $C_3$ rotation operator satisfies $r_3^3=1$, and the rotation eigenvalues are $\Pi^{(3)}_{1} = 1, \;\; \Pi^{(3)}_{2}=e^{2\pi {i}/3}, \;\; \Pi^{(3)}_{3}=e^{-2\pi {i}/3}$. The $\Pi^{(3)}_{p=1,2,3}$ eigenvalues label the $C_3$ symmetry representations of the eigenstates at the various HSPs.
If the eigenvalues within the same band differ for distinct HSPs, the energy band has a nontrivial topology.
If they are all the same, the topology is trivial. For a formal definition and evaluation of the topological indices that characterize the two phases of the non-Hermitian Kagome Hamiltonian \eqref{KagomeHamiltonian}, see Supplementary Note S2.

The non-Hermitian Hamiltonian in Eq.~\eqref{KagomeHamiltonian} has two phases that can be accessed depending on the ratio $\kappa= \kappa_\text{intra}/\kappa_\text{inter}$. For $\kappa<1$, as shown in Fig.~\ref{HighSymmetryPoints}B, the $C_3$ representations at the three HSPs $\bm{\Gamma}$, $\bm{K}$, and $\bm{K'}$ are all different within the same band, signaling a nontrivial phase.
At $\kappa=1$, when the intra- and inter-cell hopping strengths are equal, the three bands touch at the HSPs,  becoming two-fold degenerate (Fig.~\ref{HighSymmetryPoints}C). For $\kappa>1$, the symmetry representations are exchanged (compare Figs.~\ref{HighSymmetryPoints}B and \ref{HighSymmetryPoints}D), so that now all $C_3$ eigenvalues within the same band are the same, indicating a trivial phase.

\subsection{Winding numbers of counterpropagating edge bands under partial PBC}

Although the winding number of the bulk spectrum is zero, the nontrivial topological phase ($\kappa<1$), protected by line gaps that close at $\kappa=1$, supports two edge-localized bands with opposite winding numbers relative to a point gap at the origin of the complex plane.
These bands arise when we open the boundaries along one direction, say $x$ with length $L_x$, while keeping periodic boundary conditions (PBCs) along the $y$-direction, with length $L_y$.
In this configuration, the eigenstates can be localized in the $x$-direction while maintaining translation invariance along the $y$-direction, and are labeled by a one-dimensional crystal momentum $k_y \in (0,4\pi/\sqrt{3})$.

\begin{figure}
    \centering
    \includegraphics[width=0.45\textwidth]{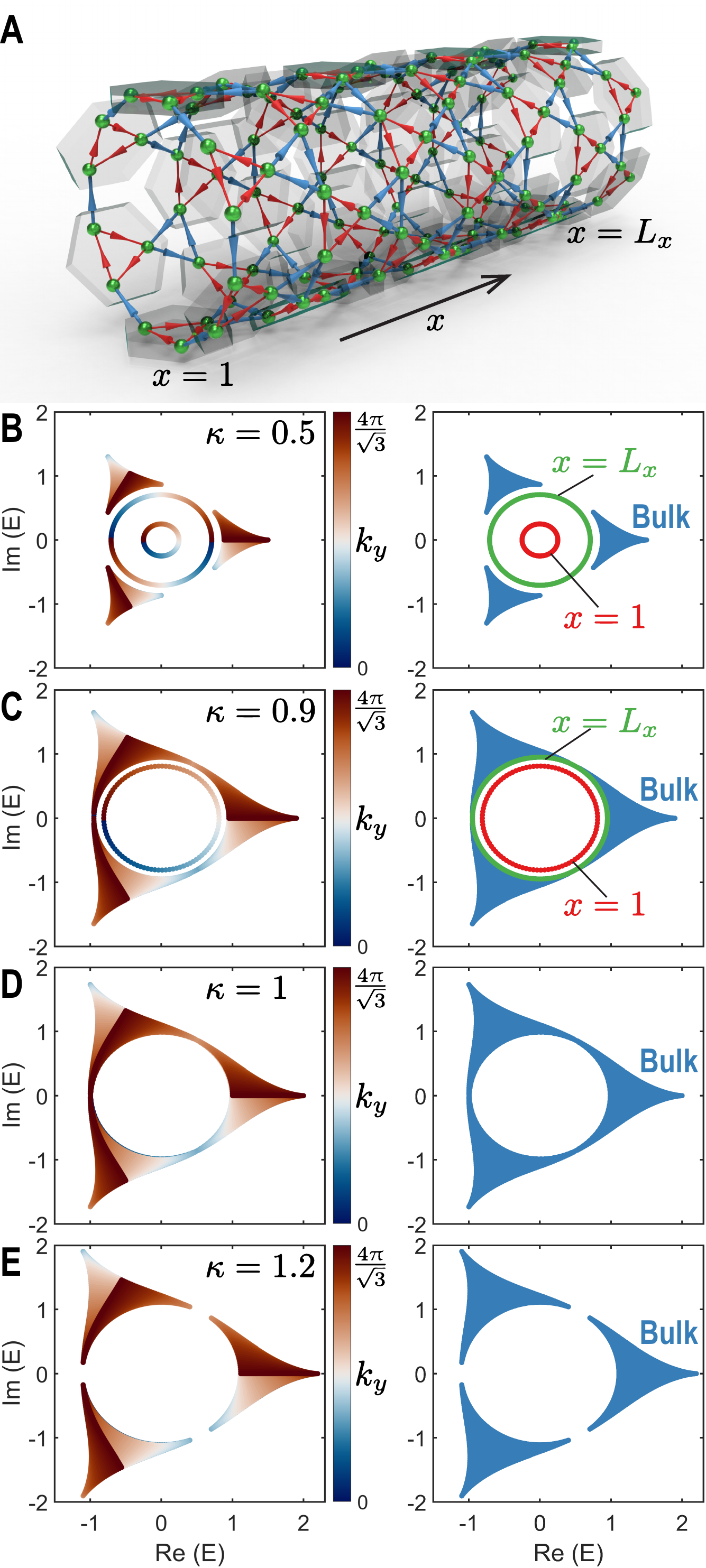}
    \caption{
        (\textbf{A}) Schematic of the non-Hermitian Kagome lattice with OBC in the $x$-direction and PBC in the $y$-direction.
        Here, the first and last unit cells in the $y$-direction of the lattice are connected to form a ring.
        (\textbf{B}--\textbf{E}) Energy spectra for (\textbf{B}) $\kappa=0.5$,
        (\textbf{C}) $\kappa=0.9$,
        (\textbf{D}) $\kappa=1$,
        and (\textbf{E}) $\kappa=1.2$, with $\kappa \equiv \kappa_\mathrm{intra} / \kappa_\mathrm{inter}$ and $\kappa_\mathrm{inter} = 1$.
        The left column shows the value of $k_y$ of the corresponding eigenstates through the color gradient, which spans the first Brillouin zone $(0,4\pi/\sqrt{3})$ in the $y$-direction.
        The right column indicates whether a given state is localized at $x=1$ (red), $x=L_x$ (green), or belongs to the bulk (blue).
    }
    \label{fig:PBC_spectrum}
\end{figure}

Figures~\ref{fig:PBC_spectrum}B and \ref{fig:PBC_spectrum}C show the energy spectrum of the non-Hermitian Kagome lattice under partial PBC (illustrated in Fig.~\ref{fig:PBC_spectrum}A) in the topological phase $\kappa<1$.
The boundary and bulk states are presented in Supplementary Fig.~S1.
In addition to the three bulk bands, there are two circles of complex energies with opposite winding in $k_y$-space.
The inner circle consists of $L_y$ eigenstates exponentially localized at the left boundary of the system ($x=1$), while the outer circle corresponds to $2L_y$ eigenstates located at the right boundary ($x=L_x$).
We can see that the right boundary has twice as many states as the left boundary.
In the inner circle, there is one state for each value of $k_y$ and as $k_y$ spans the Brillouin zone, the total set of $L_y$ states in that circumference winds counterclockwise around the point gap, resulting in a winding number $w=+1$.
On the other hand, there are two states on the outer circle for each value of $k_y$, forming two semi-circles with $L_y$ states each.
Both semi-circles half-wind clockwise around the origin as $k_y$ sweeps the Brillouin zone, resulting in a total winding number $w=-1$ for the outer circle.

Additionally, since the total winding number must always be zero by the 38-fold classification, one edge band cannot merge into the bulk while the other remains gapped.
This is particularly clear in Fig.~\ref{fig:PBC_spectrum}C, where the boundary states of the outer circle (green) ``wait'' for the coming of the inner circle (red) before overlapping with the bulk bands.
As shown in Fig.~\ref{fig:PBC_spectrum}D, both circles hybridize and disappear simultaneously at $\kappa=1$, when the line gaps close but the point gap remains open.
For $\kappa>1$, only the three trivial bulk bands remain (see Fig.~\ref{fig:PBC_spectrum}E).

Crucially, in the topological phase, the boundary at $x=1$ ($x=L_x$) acts as a 1D Hatano-Nelson chain under PBC that favors motion in the negative (positive) $y$ direction.
If we had partial PBC along the $x$ direction, the edge bands would localize at $y=1$ and $y=L_y$, favoring motion in the positive and negative $x$ directions, respectively.

\subsection{Spectrum under OBC and corner skin states}

\begin{figure}
    \centering
    \includegraphics[width = .47\textwidth]{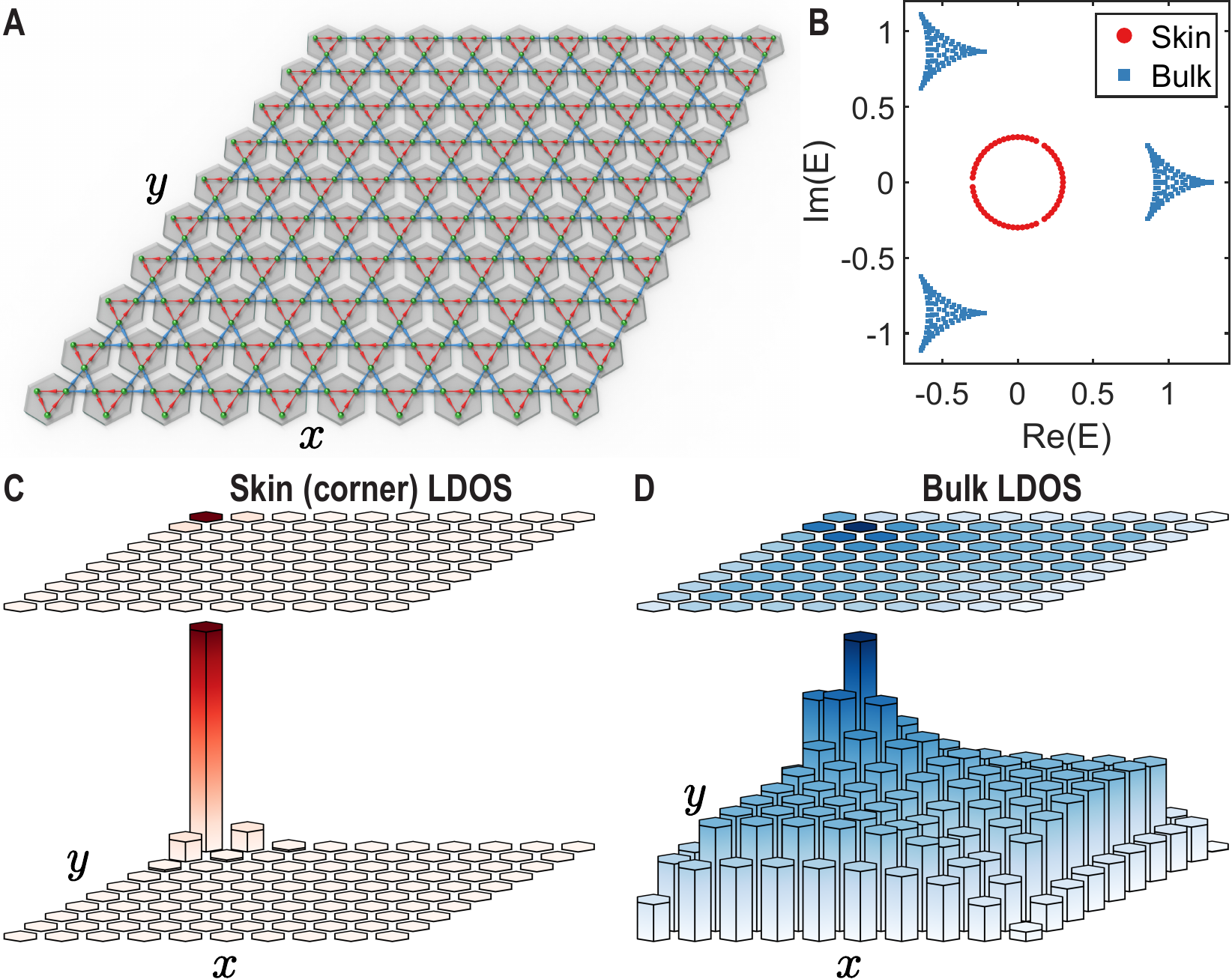}
    \caption{
        (\textbf{A}) Schematic of the non-Hermitian Kagome lattice with OBCs in both the $x$- and $y$-directions, with dimensions $L_x=L_y=10$.
        (\textbf{B}) Energy spectrum with $\kappa=0.3$, where $\kappa \equiv \kappa_\mathrm{intra} / \kappa_\mathrm{inter}$ and $\kappa_\mathrm{inter} = 1$.
        Skin (corner) states are in red while bulk states are in blue.
        (\textbf{C}) LDOS of skin (corner) states per unit cell.
        (\textbf{D}) LDOS of bulk states per unit cell.
        The LDOS distribution is normalized to the maximal value in each figure.
    }
    \label{fig:OBC}
\end{figure}

Under full open boundary conditions (OBCs), the lattice takes the shape of a parallelogram (see Fig.~\ref{fig:OBC}A) with sides $L_x=L_y=L$, allowing the eigenstates to be localized in both the $x$- and $y$-directions.
The corresponding spectrum in the nontrivial phase is displayed in Fig.~\ref{fig:OBC}B.
The modes that were previously on opposite edges forming the two rings under partial PBC are now all exponentially localized on the same region of space, as shown in Fig.~\ref{fig:OBC}C, which illustrates their local density of states (LDOS) per unit cell.
Therefore, these corner skin modes may hybridize and give rise to a single circular structure with three small line gaps, resulting in a zero winding number relative to the origin, in accordance with the constraints imposed by the model's symmetry class. Figure~\ref{fig:OBC}D shows the LDOS of the bulk bands for comparison.

Upon examining Fig.~\ref{fig:OBC}C, one can observe that the skin modes of the non-Hermitian Hamiltonian $q(\bm{k})$ in \eqref{KagomeHamiltonian} are all localized at the top left corner of the lattice.
If we were to consider its hermitian conjugate $q^{\dagger}(\bm{k})$, the skin modes would instead be localized at the bottom right corner.
This is not coincidental; their parent Hermitian Hamiltonian  \eqref{eq:Hamiltonian}, which describes a parallelogram-shaped breathing honeycomb lattice, hosts two topologically protected chiral zero modes---one at the top left corner and the other at the bottom right corner (these modes occur at corners defined by a $2\pi/3$ angle).
In general, the Hermitian-non-Hermitian correspondence predicts that each set of skin modes in $q(\bm{k})$ and $q^{\dagger}(\bm{k})$ originates from a pair of oppositely localized zero modes of $h(\bm{k})$, protected by chiral symmetry.

\section{Observation of the higher-order NHSE in non-Hermitian acoustic Kagome lattices}

\begin{figure*}[!htb]
    \centering
    \includegraphics[width = 0.95\textwidth]{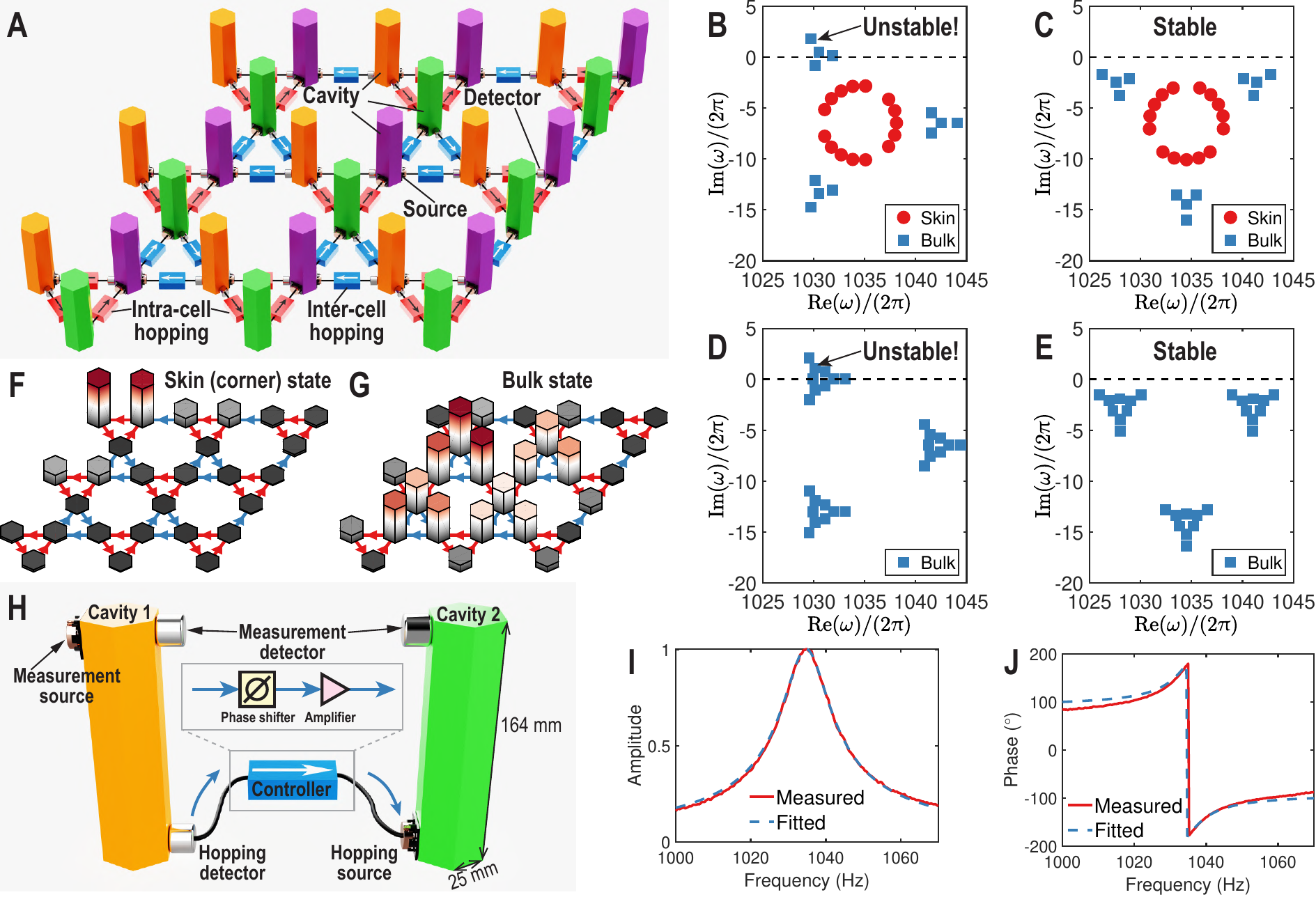}
    \caption{
        (\textbf{A}) Schematic of the non-Hermitian acoustic Kagome lattice with dimensions $L_x=L_y=3$, where the unidirectional hopping is realized by a source and detector pair, and acoustic cavities are represented by hexagonal prisms.
        (\textbf{B, C}) Complex eigenfrequencies for the nontrivial lattice obtained using the TBM with $\abs{\kappa_\mathrm{inter}} / (2\pi) = \abs{\kappa_\mathrm{intra}}/\pi = 7.43\,\mathrm{Hz}$.
        (\textbf{D, E}) Complex eigenfrequencies for the trivial lattice obtained using the TBM with $\abs{\kappa_\mathrm{intra}} / (2\pi) = \abs{\kappa_\mathrm{inter}}/\pi = 7.43\,\mathrm{Hz}$.
        The phases of both intra- and inter-cell hoppings are $\angle \kappa_\mathrm{intra} = \angle \kappa_\mathrm{inter} = 0$ in (\textbf{B, D}), while they are $\angle \kappa_\mathrm{intra} = \angle \kappa_\mathrm{inter} =-\pi/2, \pi/6$, or $5\pi/6$ in (\textbf{C, E}).
        Spatial distributions of (\textbf{F}) skin (corner) and (\textbf{G}) bulk states.
        (\textbf{H}) Schematic of the unidirectional hopping realized by a detector and a source.
        Experimentally measured and numerically fitted (\textbf{I}) amplitude and (\textbf{J}) phase response of the cross-power spectral density of the acoustic signals measured in cavities 1 and 2 shown in (\textbf{H}).
    }
    \label{fig:exp_sketch_single}
\end{figure*}

As shown in Fig.~\ref{fig:exp_sketch_single}A, the experimental platform is a non-Hermitian acoustic Kagome lattice with nonreciprocal hoppings and dimensions of  $L_x = L_y = L =3$.
The photo of the experiment setup is presented as Fig.~S2 in the Supplementary Materials.
The lattice comprises 27 3D-printed hexagonal prism-shaped acoustic cavities, each hosting the first dipole mode with a resonant frequency of 1034.5\,Hz in the vertical direction (see Appendix~\ref{app:exp_samp} for more details).
To form the lattice, the cavities are coupled through loudspeaker (source) and microphone (detector) pairs, enabling unidirectional hopping between nearest-neighbor cavities (see Fig.~\ref{fig:exp_sketch_single}H).
The eigenfrequencies in the complex plane, calculated using the TBM and parameters fitted from measurements, are presented in Figs.~\ref{fig:exp_sketch_single}B and \ref{fig:exp_sketch_single}D for the nontrivial ($\kappa = 0.5$) and trivial ($\kappa = 2$) cases, respectively.
Here, the intra- and inter-cell hoppings are real-valued with zero phase ($\kappa_\mathrm{intra},\kappa_\mathrm{inter}\in \mathbb{R}$) and are fitted through the measured signals.
In the spectrum shown in Fig.~\ref{fig:exp_sketch_single}B, there is one circumference (red dots) that hosts skin states, all localized at the top left corner (Fig.~\ref{fig:exp_sketch_single}F).
The remaining points, denoted by blue squares, host bulk states, which are demonstrated in Fig.~\ref{fig:exp_sketch_single}G.
For the trivial case shown in Fig.~\ref{fig:exp_sketch_single}D, however, only bulk states exist.

\subsection{Acoustic realization of the spectrum rotation technique }
It is important to note that when hoppings are real-valued, Figs.~\ref{fig:exp_sketch_single}B and \ref{fig:exp_sketch_single}D show that the imaginary part of some states is greater than 0, indicating pure gain at these states.
Due to the positive feedback among hoppings in the lattice, the system becomes unstable and tends to crash when any imaginary part is positive.
Therefore, to more efficiently observe the higher-order NHSE, we intentionally rotate the spectrum in the complex frequency plane by an angle, which can be either $-\pi/2,\pi/6$, or $5\pi/6$ due to $C_3$ symmetry.
This rotation ensures that all states remain below the real axis, thereby maintaining system stability, as illustrated in Figs.~\ref{fig:exp_sketch_single}C and \ref{fig:exp_sketch_single}E.
After the rotation, the imaginary parts of all eigenfrequencies are negative, ensuring system stability, while spatial distributions of skin and bulk states remain the same as shown in Figs.~\ref{fig:exp_sketch_single}F and \ref{fig:exp_sketch_single}G, respectively.
Based on the TBM, it is seen that the spectrum can be rotated by an angle of $\theta$ with respect to the on-site energy $\omega_0$ by multiplying both intra- and inter-cell hoppings with a phase term $\exp({i}\theta)$, i.e., $\kappa_\mathrm{intra} \to \kappa_\mathrm{intra} \exp({i}\theta)$ and $\kappa_\mathrm{inter} \to \kappa_\mathrm{inter} \exp({i}\theta)$, while their ratio remains the same, $\kappa = \kappa_\mathrm{intra} / \kappa_\mathrm{inter}$.

To validate the spectrum rotation technique in experiments, we start with a simple coupled acoustic cavity system as shown in Fig.~\ref{fig:exp_sketch_single}H.
The unidirectional hopping is realized by active acoustic components, including a microphone (hopping detector), a loudspeaker (hopping source), an amplifier, and a phase shifter.
The microphone is placed in cavity 1 to detect the acoustic pressure inside this cavity.
The phase of the detected acoustic pressure is adjusted, the signal is amplified, and then emitted by the loudspeaker placed in cavity 2.
The tight-binding Hamiltonian of this system is
\begin{equation}
    H_2 =  \mqty(\omega_0 & 0 \\ \kappa_0 & \omega_0 )
    .
    \label{eq:cacs_uni_hamiltonian}
\end{equation}
Here, $\kappa_0\in\mathbb{C}$ is the unidirectional hopping from cavity 1 to cavity 2, which is complex-valued and can be either intra- or inter-cell hopping.
The real and imaginary parts of the on-site energy, $\omega_0$, represent the first resonant frequency and the intrinsic loss of a single cavity, respectively.
By exciting a single cavity with a loudspeaker and recording the signal using a microphone, the on-site frequency can be fitted to approximately $\omega_0/(2\pi) = 1034.5\,\mathrm{Hz} -6.37 i\,\mathrm{Hz}$ (see Appendix~\ref{app:exp_samp} for more details).

To determine both the amplitude and the phase of the unidirectional hopping $\kappa_0$ in Eq.~\eqref{eq:cacs_uni_hamiltonian}, we excite at cavity 1 using a loudspeaker (measurement source in Fig.~\ref{fig:exp_sketch_single}H) and measure the acoustic pressure in both cavities using two microphones.
The cross-power spectral density of these two measured pressure signals can be expressed as $G(\omega) = \psi_2(\omega)/\psi_1(\omega) = \kappa_0 / (\omega - \omega_0)$, where $\omega$ is the excitation frequency, and $\psi_i$ represents the acoustic pressure in cavity $i$ with $i=1,2$ (see Supplementary Note S3 for details).
It is clear that the spectral response of $\abs{G(\omega)}$ has a single peak at $\omega = \Re(\omega_0)$.
Furthermore, when $\omega = \Re(\omega_0)$, the phase of $G$ and $\kappa_0$ follows the relation
\begin{equation}
    \angle G(\Re(\omega_0)) = \angle \kappa_0 - \frac{\pi}{ 2}.
    \label{eq:angle}
\end{equation}
This relation can be used to adjust the phase of $\kappa_0$ by observing the phase response of $G(\omega)$.
In the experiment, we set $\kappa_0$ to $-\pi/2$, indicating a counterclockwise rotation of the spectrum by an angle of $-\pi/2$, producing rotated spectrum shown in Figs.~\ref{fig:exp_sketch_single}C and \ref{fig:exp_sketch_single}D.
Consequently, the phase response of $G(\omega)$ at $\omega=\Re(\omega_0)$ should be $-\pi$ according to Eq.~\eqref{eq:angle}.
Although analog circuits can be designed to shift the phase of the hopping \cite{Wang2023ExtendedTopologicalMode, Zhang2023ObservationAcousticNonHermitian, Zhang2023ExperimentalCharacterizationThreeband, Chen2023SoundNonreciprocityBased, Chen2024TransientLogicOperations, Chen2024RobustTemporalAdiabatic}, they are challenging and time-consuming to adjust precisely for a large lattice.
In this work, digital fractional delay filters based on the Lagrange interpolation \cite{Laakso1996SplittingUnitDelay, Smith2010PhysicalAudioSignal} are implemented by a custom-made real-time controller to shift the phase (see Appendixes~\ref{app:impl_uni_hop} and \ref{sec:stability} for more details).
This digital filter approach offers much more precise control of the phase and greater flexibility, especially for adjusting hopping strengths for different topological phases.
Figures~\ref{fig:exp_sketch_single}I and \ref{fig:exp_sketch_single}J show the amplitude and phase response of $G(\omega)$, where measured results agree well with the fitted ones, validating our technique.

\subsection{Observation of the higher-order NHSE}
\begin{figure*}[!htb]
    \centering
    \includegraphics[width = 0.99\textwidth]{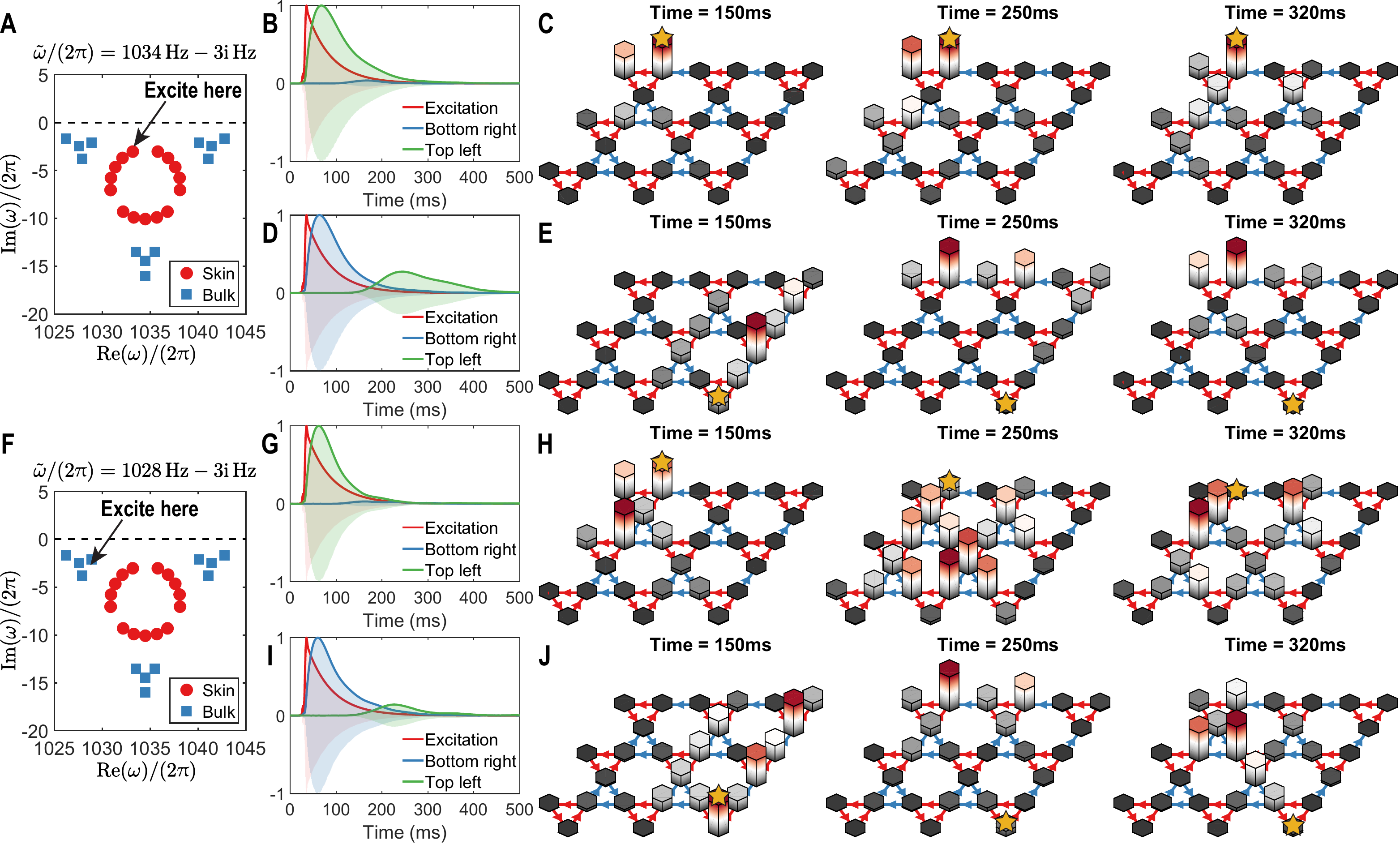}
    \caption{
    Experimental demonstration in a nontrivial Kagome lattice with dimensions $L_x=L_y=3$ and hopping parameters $\kappa=0.5$ and  ${\kappa_\mathrm{inter}} / (2\pi) = {\kappa_\mathrm{intra}}/\pi = 7.43
        e^{-i\pi/2} \,\mathrm{Hz}$.
    (\textbf{A}, \textbf{F}) The complex eigenfrequencies showing the CFE at (\textbf{A}) $\tilde{\omega}/(2\pi) = 1034\,\mathrm{Hz}-3{i}\,\mathrm{Hz}$ and (\textbf{F}) $\tilde{\omega}/(2\pi) = 1028\,\mathrm{Hz}-3{i}\,\mathrm{Hz}$.
    The lattice is excited by a loudspeaker (denoted by a yellow star
    \begin{tikzpicture}[scale = .16]
        \fill[fill={rgb,255:red,236; green,176; blue,32}, draw=black] (0,1) -- (0.2245,0.309) -- (0.951,0.309) -- (0.3633,-0.118) -- (0.5878,-0.809) -- (0, -0.382) -- (-0.5878,-0.809) -- (-0.3633,-0.118) -- (-0.951,0.309) -- (-0.2245,0.309) -- cycle;
    \end{tikzpicture}
    ) at a cavity in the (\textbf{B, C, G, H}) top left and (\textbf{D, E, I, J}) bottom right unit cells.
    (\textbf{B, D, G, I}) The excitation signal and signals measured at the cavity in the top left and bottom right unit cells as a function of time.
    (\textbf{C, E, H, J}) Acoustic energy distributions measured at 150\,{ms}, 250\,ms, and 320\,ms.
    }
    \label{fig:exp_result_nontrivial}
\end{figure*}
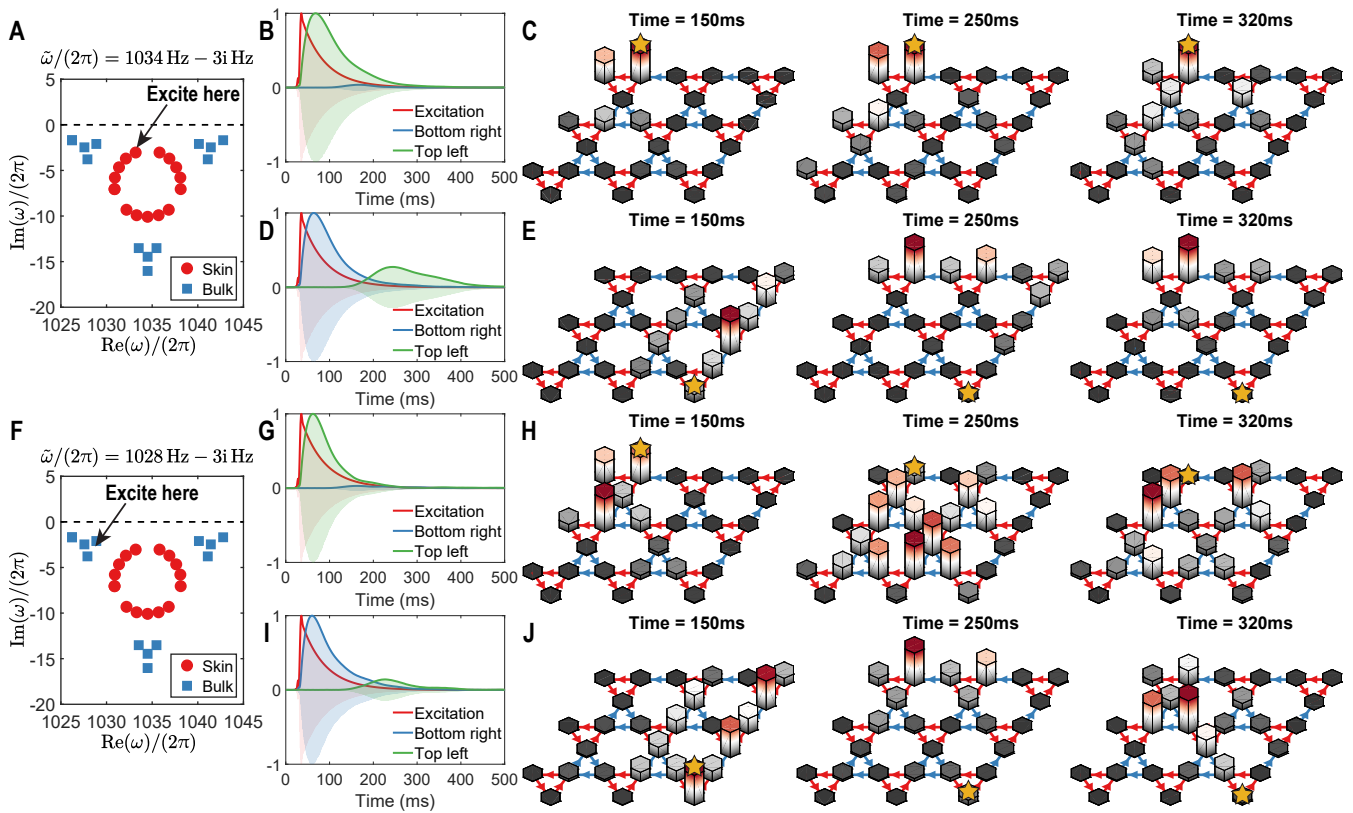

To observe the higher-order NHSE, we excite the lattice using a loudspeaker at one cavity (denoted by the star in Fig.~\ref{fig:exp_result_nontrivial}) and measure the acoustic pressure signal at every cavity using a microphone (see Appendix~\ref{sec:complex} for more details).
It is important to note that, to ensure system stability, the imaginary part of the frequencies of the skin states must be negative (indicating physical loss), as shown in Fig.~\ref{fig:exp_result_nontrivial}A.
Consequently, the observation of the NHSE would be obscured by the loss \cite{Zhang2021ObservationHigherorderNonHermitian, Gu2022TransientNonHermitianSkin}.
To address this issue, we adopt the CFE technique \cite{Li2020VirtualParityTimeSymmetry, Kim2022BoundsLightScattering, Kim2023LossCompensationSuperresolution, Gu2022TransientNonHermitianSkin, Guan2023OvercomingLossesSuperlenses, Guan2024CompensatingLossesPolariton, Zeng2024SynthesizedComplexfrequencyExcitation} to observe the NHSE.
The CFE technique involves exciting with an exponentially decayed sinusoidal signal in the time domain, where the decay in the signal amplitude introduces an effective virtual gain to compensate for the loss (see Appendix~\ref{sec:complex} for details).
For comparison, the results with traditional real-frequency excitation are presented in Figs.~S5 and S6 in Supplementary Materials.

\begin{figure*}[!htb]
    \centering
    \includegraphics[width = 0.99\textwidth]{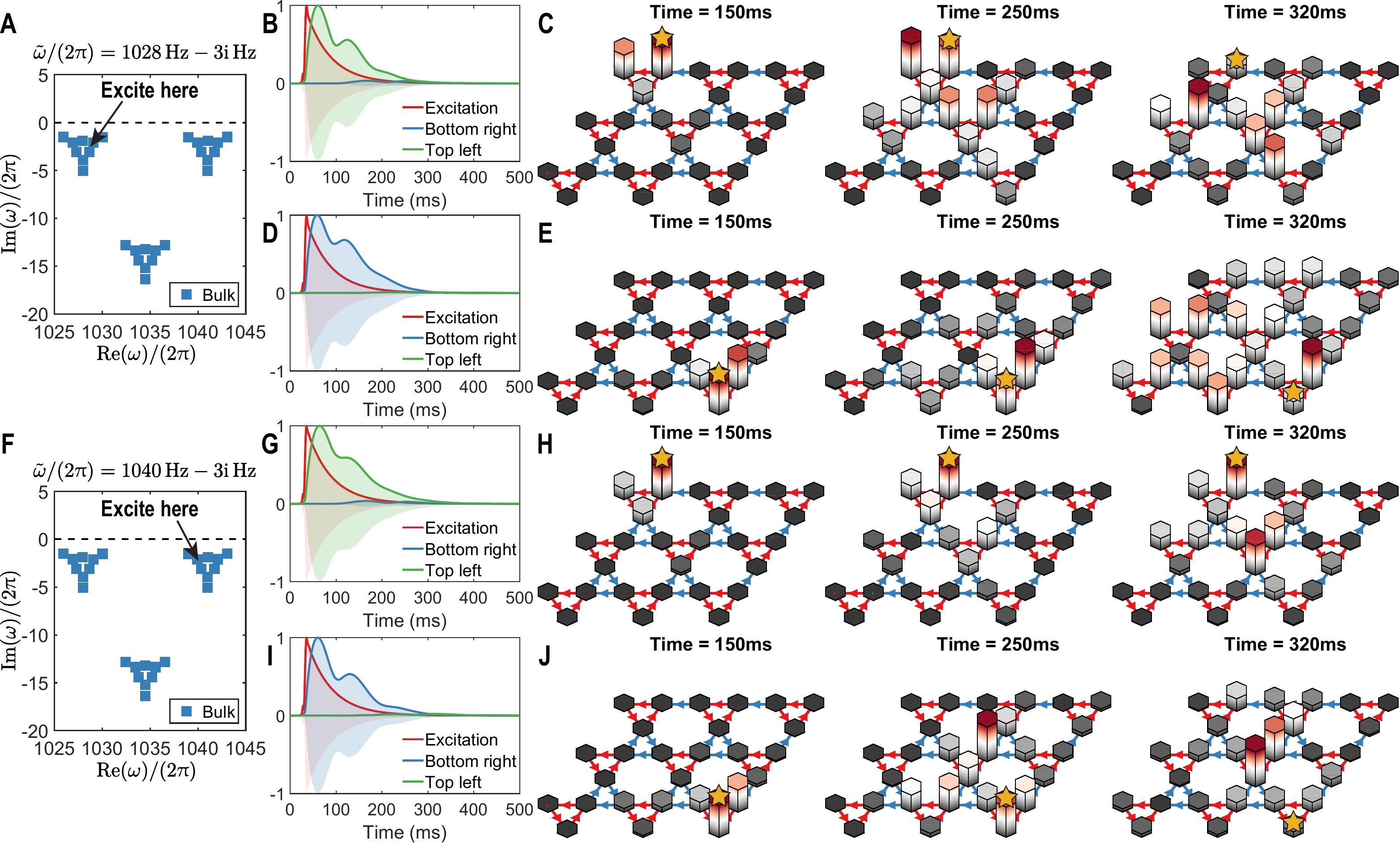}
    \caption{
    Experimental demonstration in a trivial Kagome lattice with dimensions $L_x=L_y=3$ and hopping parameters $\kappa=2$ and  ${\kappa_\mathrm{intra}} / (2\pi) = {\kappa_\mathrm{inter}}/\pi = 7.43
        e^{-i\pi/2} \,\mathrm{Hz}$.
    (\textbf{A}, \textbf{F}) The complex eigenfrequencies showing the CFE at (\textbf{A}) $\tilde{\omega}/(2\pi)=1034\,\mathrm{Hz}-3{i}\,\mathrm{Hz}$ and (\textbf{F}) $\tilde{\omega}/(2\pi)=1028\,\mathrm{Hz}-3{i}\,\mathrm{Hz}$.
    The lattice is excited by a loudspeaker (denoted by a yellow star
    \begin{tikzpicture}[scale = .16]
        \fill[fill={rgb,255:red,236; green,176; blue,32}, draw=black] (0,1) -- (0.2245,0.309) -- (0.951,0.309) -- (0.3633,-0.118) -- (0.5878,-0.809) -- (0, -0.382) -- (-0.5878,-0.809) -- (-0.3633,-0.118) -- (-0.951,0.309) -- (-0.2245,0.309) -- cycle;
    \end{tikzpicture}
    ) at a cavity in the (\textbf{B, C, G, H}) top left and (\textbf{D, E, I, J}) bottom right unit cells.
    (\textbf{B, D, G, I}) The excitation signal and signals measured at the cavity in the top left and bottom right unit cells as a function of time.
    (\textbf{C, E, H, J}) Acoustic energy distributions measured at 150\,{ms}, 250\,ms, and 320\,ms.
    }
    \label{fig:exp_result_trivial}
\end{figure*}
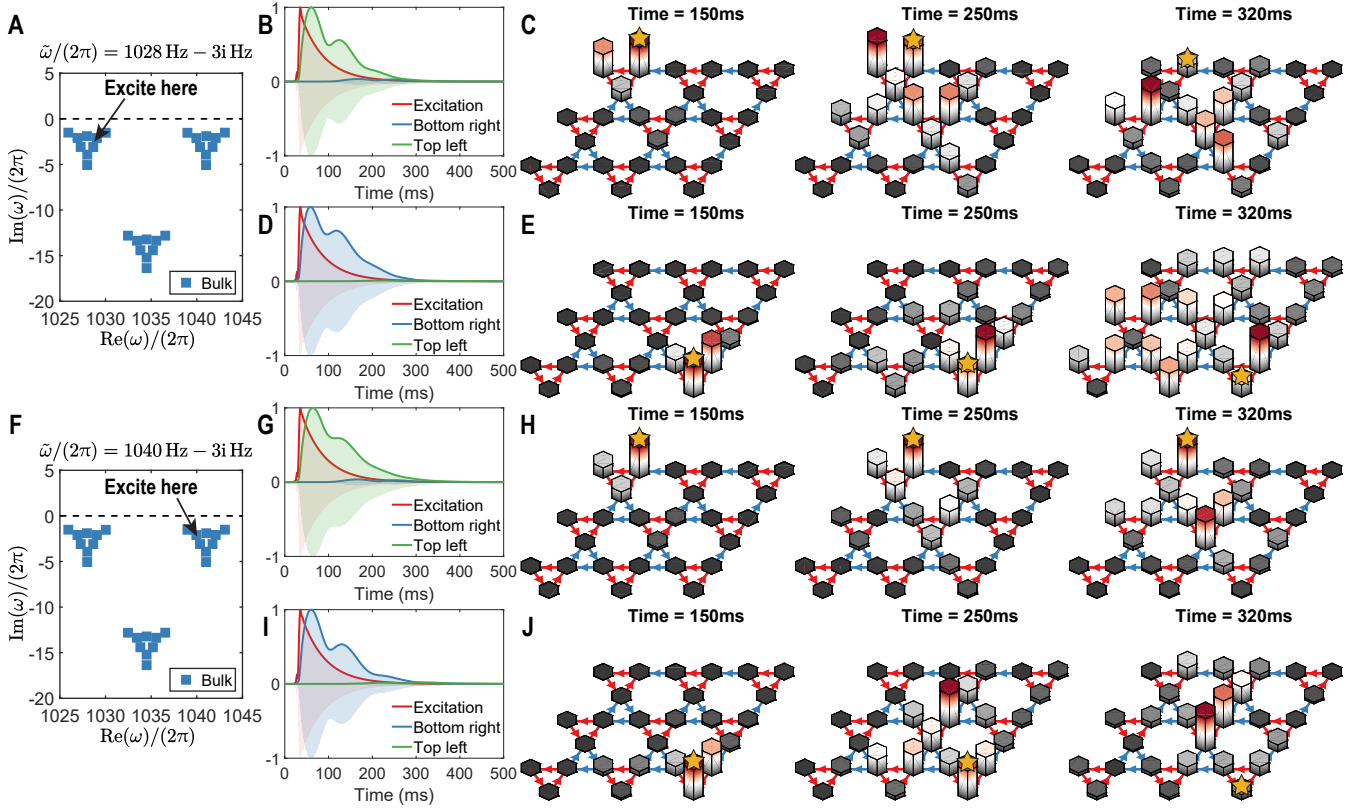

For the nontrivial lattice with $\kappa = 0.5$ as shown in Figs.~\ref{fig:exp_result_nontrivial}A and \ref{fig:exp_result_nontrivial}F, we implement the CFE at $\tilde{\omega} /(2\pi) = 1034\,\mathrm{Hz}-3i\,\mathrm{Hz}$ and $1028\,\mathrm{Hz}-3i\,\mathrm{Hz}$, respectively, to observe the NHSE and bulk states, respectively.
Figure~\ref{fig:exp_result_nontrivial}B shows the time evolution of the measured excitation signal and acoustic signals in cavities located in bottom right and top left unit cells.
As observed in both Figs.~\ref{fig:exp_result_nontrivial}B and \ref{fig:exp_result_nontrivial}C, the acoustic energy remains localized at the top left corner most of the time, as expected in Fig.~\ref{fig:exp_sketch_single}F.
When the source is excited at the bottom right unit cell, the furthest from the top left corner, the time-evolved signals shown in Fig.~\ref{fig:exp_result_nontrivial}D and the spatial distributions of the acoustic energy in Fig.~\ref{fig:exp_result_nontrivial}E demonstrates that the system reaches a stable quasi-stationary stage after around 250\,ms, and the acoustic energy is localized at the top left unit cell, compatible with the transport expected for the NHSE.
In this context, the transport response refers to the localization of wave energy to a specific corner, regardless of the excitation location.
For comparison, as shown in Fig.~\ref{fig:exp_result_nontrivial}F, we also measure the time-evolved signals at $\tilde{\omega} /(2\pi)  = 1028\,\mathrm{Hz}-3i\,\mathrm{Hz}$, which is expected to host bulk states.
The measured results presented in Figs.~\ref{fig:exp_result_nontrivial}G--\ref{fig:exp_result_nontrivial}J clearly demonstrate that bulk states are excited instead of corner ones.
Detailed time evolution of the acoustic energy fields, as well as the excitation at $\tilde{\omega} /(2\pi)  = 1040\,\mathrm{Hz}-3i\,\mathrm{Hz}$ hosting bulk states at the other band, are provided as movies and Fig.~S3 in Supplementary Materials.
For better comparison, we also measure the trivial lattice with $\kappa = 2$ and results are summarized in Fig.~\ref{fig:exp_result_trivial}.
The extended data for the trivial lattice are also provided as movies and Fig.~S4 in Supplementary Materials.
There are only three bulk bands shown in the complex spectrum in Figs.~\ref{fig:exp_result_trivial}A.
We then launch the CFE at $1028\,\mathrm{Hz}-3i\,\mathrm{Hz}$ and $1040\,\mathrm{Hz}-3i\,\mathrm{Hz}$ in Figs.~\ref{fig:exp_result_trivial}A and \ref{fig:exp_result_trivial}F, respectively.
The time-evolved signals and the spatial distributions of acoustic energy in Fig.~\ref{fig:exp_result_trivial} demonstrate the bulk states as expected.


\section{Discussion}
Through a Hermitian-non-Hermitian correspondence, we identify a recipe for constructing lattices with higher-order NHSE from a HOTI protected by chiral symmetry.
Specifically, we construct a non-Hermitian Kagome lattice protected by $C_3$ symmetry from a Hermitian breathing honeycomb lattice.
We also report the experimental observation of the higher-order NHSE in this non-Hermitian acoustic Kagome lattice, where nonreciprocity is manifested by the unidirectional amplification of acoustic signals.
To minimize the loss of skin states while maintaining system stability, we intentionally rotate the complex spectrum by implementing complex hoppings.
Additionally, we utilize the CFE approach, which introduces an effective virtual gain to compensate for the intrinsic loss in the lattice. These techniques allow us to observe the localization of acoustic energy to a corner far away from the excitation location in the topological phase with higher-order NHSE.
Our work paves the way for a deeper understanding of higher-order NHSE and potentially innovative applications rooted in acoustic nonreciprocity, such as robust waveguides, nonreciprocal amplifiers \cite{Shao2020NonreciprocalTransmissionMicrowave}, and so on \cite{Shah2024ColloquiumTopologicallyProtected, Rasmussen2021AcousticNonreciprocity}.

\begin{acknowledgments}
    W. A. B. thanks the support of startup funds from Emory University.
    Y. J. thanks the support of startup funds from Penn State University and NSF awards 2039463 and 1951221.
    J. L. gratefully acknowledges the financial support by National Natural Science Foundation of China (12274221).
\end{acknowledgments}

\appendix*

\section{Experimental details}

\subsection{
    \label{app:exp_samp}
    Experimental samples }
The acoustic cavities used in this work were fabricated using 3D printers with a tolerance of 0.2\,mm or within 0.3\%.
The materials are LEDO 6060 photosensitive resin, which can be considered as acoustically rigid for airborne sound.
The printed samples have a thickness of 6\,mm.
As shown in Fig.~\ref{fig:exp_sketch_single}, all acoustic cavities used in experiments are hexagonal prisms with an interior height of $l = 164\,\mathrm{mm}$ and a side length of $25\,\mathrm{mm}$.
Each acoustic cavity hosts a resonant mode at a frequency of $c_0/(2l)\approx 1045.7\,\mathrm{Hz}$, where $c_0=343\,\mathrm{m/s}$ is the airborne sound speed.
In the experiments, a loudspeaker (source) is used to excite a single acoustic cavity, and a microphone (detector) is used to measure the acoustic pressure in that cavity.
By using the Green's function for a single mode \cite{Li2023EigenvalueKnotsTheir}, $\psi(\omega) = -\Im(\omega_0)/(\omega - \omega_0)$, we retrieve the on-site orbit as $\omega_0/(2\pi)=1034.5\,\mathrm{Hz} - 6.37i\,\mathrm{Hz}$, where $\omega$ is the excitation frequency.

\subsection{
    \label{app:impl_uni_hop}
    Implementation of unidirectional hoppings}
As shown in Fig.~\ref{fig:exp_sketch_single}H, each unidirectional hopping comprises a loudspeaker (CDS-25148-L100, CUI Devices), a microphone (BOB-19389, SparkFun Electronics), and an audio amplifier (EK1236 module equipped with an onboard LM386 chip).
The loudspeaker and microphone are placed at the bottom of each cavity.
We design and develop a custom-made digital real-time controller to implement the unidirectional hoppings in our model.
This allows more flexible control and scalability relative to existing analog amplifier and phase shifter-based approaches \cite{Zhang2021AcousticNonHermitianSkin, Liu2022ExperimentalRealizationWeyl, Zhang2023ObservationAcousticNonHermitian, Wang2023ExtendedTopologicalMode, Wang2022NonHermitianMorphingTopological}, which become necessary, particularly to capture transport phenomena across topological phase transitions.
The microphone captures the acoustic pressure signal, which is then processed by the controller set to a sampling rate of 80\,kHz.
In this work, digital fractional delay filters based on the Lagrange interpolation \cite{Laakso1996SplittingUnitDelay, Smith2010PhysicalAudioSignal} are implemented by the controller to shift the phase.

The custom-made digital controller consists of three parts: the core board, the motherboard, and
the input/output (IO) board.
The core board integrates a field programmable gate array (FPGA, XC7K325T, Xilinx) and a digital signal processor (DSP, TMS320C6678, Texas Instruments).
The motherboard provides power, and communication buses to the core and IO boards.
The IO board acquires and generates real-time signals.
Each IO board supports 16-channel analog inputs and outputs.
The input channel employs analog-to-digital converters (ADC, ADC7606B, ADI), and
the output channel adopts digital-to-analog converters (DAC, DAC8568, Texas Instruments).

\subsection{
    \label{sec:stability}
    Stability control and the energy-spectrum rotation}
It is noted that feedback within the entire lattice can make the system unstable at acoustic resonant modes.
The physical mechanism behind this is that any states with pure gain can cause the system to crash, inducing unstable positive feedback similar to howling sound effects.
As demonstrated in Fig.~\ref{fig:exp_sketch_single}, stability around the working resonant mode at $\Re(\omega_0)$ can be enhanced by rotating the complex energy spectrum to ensure all imaginary parts are less than zero.
While reducing the loss of skin states at the working resonant mode $\Re(\omega_0)$, the higher-order harmonic resonant modes of the acoustic cavities, $j\Re(\omega_0)$, where $j=2,3,...$, can also introduce stability issue.
To address this issue, we use digital notch filters (second-order infinite impulse response filters) implemented by the controller to mitigate the 2nd, 3rd, 4th, and 5th order resonances at 2069\,Hz, 3103.5\,Hz, 4138\,Hz, and 5172.5\,Hz, respectively.
The $Q$ factor of the notch filter is set to 10 in experiments.

\subsection{
    \label{sec:complex}
    Complex-frequency excitation and measurements}
For the purpose of ensuring the imaginary parts of all energies are negative, there is a small background loss across the lattice, as shown in Fig.~\ref{fig:exp_sketch_single}.
The loss may obscure the higher-order NHSE to a certain extent \cite{Zhang2021ObservationHigherorderNonHermitian, Gu2022TransientNonHermitianSkin}.
To more clearly observe the higher-order NHSE, we apply the CFE technique.
To illustrate the basic idea of the CFE, consider a simple lossy 1D wave propagation model with a uniform loss $\gamma_0 >0$, where the signal decays spatially according to $e^{-\gamma_0z}$, with $z>0$ being the spatial coordinate.
Suppose a source is excited at $z=0$ with a signal $s(t)$ at time $t$; the signal is then expressed in the temporal-spatial domain as $s\qty(t - z/c_\mathrm{p}) e^{-\gamma_0z}$, where $c_\mathrm{p}>0$ is the wave speed.
If the excitation signal oscillates at a complex frequency $\tilde{\omega} = \Re(\tilde{\omega}) + i \Im(\tilde{\omega})$, it takes the form $s(t) = e^{-i \tilde{\omega}t}$.
Consequently, the wave field is expressed as $e^{-i\Re(\tilde{\omega}) (t-z/c_\mathrm{p})} e^{\Im (\tilde{\omega}) t - (\gamma_0 +\Im(\tilde{\omega}) / c_\mathrm{p})z}$ with its magnitude of $e^{\Im (\tilde{\omega}) t - (\gamma_0 +\Im(\tilde{\omega}) / c_\mathrm{p})z}$.
Clearly, when $\Im(\tilde{\omega}) = -\gamma_0 c_\mathrm{p}<0$, the magnitude of the wave field becomes independent of the spatial coordinate $z$.
A negative imaginary part of $\tilde{\omega}$, i.e., $\Im(\tilde{\omega})<0$, represents a decaying signal in the time domain, effectively providing a virtual gain to compensate for the spatial loss in the medium.

In experiments, the CFE signal is a two-second sinusoidal signal with an exponential decaying rate in amplitude, with expression $s(t) = \sin[\Re(\tilde\omega) t] {e}^{\Im(\tilde\omega) t} [u(t) - u(t-2)]$, where $u(t)$ is the Heaviside function.
All signals are generated, acquired, and post-processed by a generator module (B\&K Type 3160-A-042 equipped with a LAN-XI panel Type UA-2102-042).
The excitation signal is generated from the generator module and drives a loudspeaker (CMS-15118D-L100, CUI Devices) at the top of a cavity.
The coustic pressure is measured by 1/4-inch condenser microphones (B\&K Type 4187 equipped with a preamplifier Type 2670) in the receiver cavity.
The acoustic spectral response was obtained using the ``H1 frequency response'' settings in the B\&K PULSE LabShop software.

\nocite{*}

%

\end{document}